\documentstyle[preprint,tighten,aps,epsfig]{revtex}
\begin{document}


\draft
\title{Cosmic microwave background and parametric resonance in reheating}
\author{A. B. Henriques}
\address{Departamento de Fisica/CENTRA,Instituto Superior Tecnico,
1096 Lisbon, Portugal}
\author{R. G. Moorhouse}
\address{Department of Physics and Astronomy,University of Glasgow,
Glasgow G12 8QQ, U.K.}
\date{\today}
\maketitle
\begin{abstract}
The variation of the perturbative 3-curvature  parameter, $\zeta$,
 is investigated in the period of reheating after inflation. The two-field 
 model used has the inflaton, with an  extra scalar field coupled to it,
 and non-linear effects of both fields are included as well as a slow 
 decay mechanism into the hydrodynamic fluid of the radiation era. 
 Changes in $\zeta$ occur and  persist into the succeeding cosmic eras 
 to influence the generation  of the cosmic microwave background 
 fluctuations. With appropriate  coupling the extra field rises 
 from almost zero at the end of inflation to parametric resonance which
 is important in giving  ultimate changes to $\zeta$. Inflaton particles 
 are always produced in important quantities and their interaction with 
 the extra field may limit its parametric resonance. When the field 
 coupling is such that parametric resonance is small or absent then
 non-linear effects of the produced inflaton particles by themselves
 also give ultimate changes in $\zeta$, though these are small.  
      
\end{abstract} 
\section{INTRODUCTION}\label{INTRO}
     In the cosmic inflationary scenario parametric excitation of extra
fields may mediate the transition of the inflaton field into a cosmic 
radiation era of a gas of relativistic particles
\cite{TRA,KLS1,etc,KLS2,K,KT,Boy,Fin,FB}. There has been discussion of the 
possible role of non-linear effects in such transitions 
\cite{HM1,HM2,LLMW,BassV3,JS}; depending on their magnitude such effects 
 may be important in connecting the quantum excitations of the inflationary 
 era inflaton field to the cosmic microwave background radiation 
 fluctuations (CMBRF). When non-linear effects are not important this
 connection has often been made rather directly, for example by using
 the time-invariant properties of a parameter $\zeta$, associated with spatial 
 curvature perturbations \cite{BST,LYTH,MUK}, for wavelengths appropriate 
 to the CMBRF. To study such non-linear reheating effects on the CMBRF we 
 calculate using the simplest theory as a test-bed with just
 one additional scalar field subject to parametric resonance \cite{HM1,HM2}.   

 Thus we have a model with an inflaton field
$\varphi=\varphi_0+\varphi_1$, where $\varphi_0$ denotes the 'classical'
part of the inflaton field and $\varphi_1$ the part arising from 
primordial quantum perturbations of $\varphi$. We use the perturbed FRW
metric in longitudinal gauge which appears with the perturbing fields 
$\psi$ and $\Phi$ as
\begin{equation}\label{eqI1}
ds^2 = a(\tau)^2(1+2\Phi)d\tau^2-a(\tau)^2(1-2\psi)\delta_{ij}dx^idx^j. 
\end{equation}
In the usual case of absence of space-space off-diagonal elements of the 
stress-energy tensor $\psi$ and $\Phi$ are equal.
The one additional field $\chi$ is taken to have the well-used interaction
term $\frac{1}{2}g^2\varphi^2\chi^2$ \cite{TRA,KLS1,etc,KLS2,K}. Persuasive
arguments have been given \cite{LLMW,BassV3,JS} that $\chi$ is strongly suppressed
before reheating begins. It is then reasonable to treat $\chi$ as a field 
arising from its own quantum fluctuations without any primordial classical 
part. 
We thus have 3 initial quantum fields $\psi$, $\varphi_1$ and  $\chi$.
 The equations of motion imply that in the canonical decomposition of quantum 
 fields into momentum component creation and annihilation operators, the 
 operators for $\psi$ and $\varphi_1$ are the same and those for $\chi$ are
independent.

As shown by Polarski and Starobinsky \cite{PS} the quantum fields decohere into
classical stochastic fields which nevertheless keep important properties of 
the initial quantum fields; for instance the ensemble averaging of the products 
of stochastic fields give similar results to the vacuum expectation value of 
 the corresponding quantum fields. Since we cannot realistically invoke a
 non-perturbative and exact quantum mechanical treatment such averages necessarily
 arise in  our treatment of the equations of motion; for example from   
 $\frac{1}{2}g^2\varphi^2\chi^2$ a term of the form $g^2\varphi_0\chi^2$ 
 appears in the equation of motion of the classical homogeneous part 
 $\varphi_0$ of the inflaton requiring for consistency the introduction of the
 ensemble average, $<\chi^2>$, of $\chi^2$. 

This feature is important, for example in the case of the homogeneous Einstein 
equations where the contributions due to $\varphi_1$ and $\chi$
during the reheating period may become the dominant ones. Also in the 
 non-homogneous Einstein equations  $\varphi_1$ cannot 
 be treated as a perturbation but should be viewed as the non-homogeneous part
of the field $\varphi$. On the other hand the field $\psi(x,t)$
will be treated as a perturbation and in our numerical simulations checks will
always be made to ensure that $\psi(x,t)$ is always much smaller than unity.
The justification of this procedure has been discussed previously \cite{HM1,HM2}.

  Our model also incorporates true 
heating through an elementary decay mechanism, with decay constants 
 $\Gamma_{\varphi}$ and $\Gamma_{\chi}$, of the scalar fields into a hydrodynamical 
relativistic (radiation) fluid. The system thermalizes well 
 with $\psi$ becoming approximately constant as expected from analytic calculations.
 We calculate, using well-known procedures, 
the cosmic microwave background fluctuations starting from the 
 values of  $\psi$ or $\zeta$ immediately after thermalization thus incorporating 
 any changes that may have occured in the transition from inflation to the 
 radiation era.  
             
 We shall see that we have to deal with two different regimes depending on 
 the relative values of the particle or field energy densities attained
 in the transition. On the one hand we have the classical inflaton field,
 $\varphi_0$, whose density declines plus the inflaton particles field,
 $\varphi_1$, whose density increases; these densities become close to each 
 other and stay so until they thermalize. On the other hand we have the
 $\chi$ density which at some time in the transition increases rapidly 
 from almost zero to some non-negligable value. One regime, A, is that
 where  the $\chi$ density stays below the other densities;
 the second regime, B, which is induced by larger couplings than those of
 A is where the the $\chi$  density approaches rivalry 
 with the others. Both regimes can be induced by a large number of different
 values of the parameters $g^2$, the coupling strength, and of
 $\Gamma_{\varphi}$ and $\Gamma_{\chi}$ which vary the rapidity 
 of thermalization. In both regimes $\zeta$ changes but in A, where the $\chi$
 density remains comparatively modest, the change is constant and smaller 
 while in B  the change factor varies.

We know from analytic argument that when there are non-linear effects of the 
 fields the usual theorems showing $\zeta$ to be constant fail, for example
 as shown in section \ref{subsecIIIB} below. Because of the complexity of
 the equations we do need numerical computation to verify and quantify 
 the change in $\zeta$ and any importance it may have for calculation of
 the CMBRF. This paper attempts to do so for a particular model of 
 preheating and reheating.

The paper has been organised as follows. In the next section we precisely 
define the model for the reheating period and give the resulting equations 
of motion. The initial conditions arising from the inflationary period are also
 given. We call attention to the fact that despite the initial extreme suppression 
and smallness of $\chi$ \cite{LLMW,BassV3,JS} this field still develops an 
important resonating behaviour. The theory can be formulated in different 
ways because of the redundancy of the many equations of motion and 
 some small such changes have been made with respect to our 
previous papers \cite{HM1,HM2} with the purpose of increasing the accuracy of the 
 numerical computations. In section III we discuss the results of the computations
 and in section IV we make the connection with the cosmic microwave background 
 fluctuations.

\section{THE REHEATING EQUATIONS}
\label{secII}

\subsection{The Equations of Motion}
\label{subsecIIA}
The scalar field part of the model is specified by the Lagrangian
\begin{equation}\label{eqIIA1}
L=\int d^4x\sqrt{-g}\lbrack\frac{1}{2}\varphi^{,\alpha}\varphi_{,\alpha}+
\frac{1}{2}\chi^{,\alpha}\chi_{,\alpha}-V(\varphi)-V(\chi)-
V_{int}(\varphi,\chi)\rbrack 
\end{equation}
where $\varphi=\varphi_0 + \varphi_1$..
We take the field potentials during reheating to be
\begin{equation}\label{eqIIA2}
 V(\varphi)=\frac{1}{2}m^2\varphi^2 ; V(\chi)=\frac{1}{2}M^2\chi^2; 
  V_{int}(\varphi,\chi) = \frac{1}{2} g^2 \varphi^2 \chi^2. 
\end{equation} 
 The perturbed longitudinal gauge metric is given by Eq.(\ref{eqI1}),
 the dynamics of the evolution of the scale $a(\tau)$ by the Einstein
homogeneous equation (\ref{eqIIA11}) below. It is characteristic of our 
treatment that we always include the metric perturbations. As in our
physical system there are no space-space off-diagonal elements of the 
stress-energy tensor \footnote{The ensemble  averaging of
 non linear terms, described 
below and more fully in ref.\cite{HM1}, suppresses any such terms which 
might otherwise be present in our non-linear theory. Corrections
 to the result $\Phi=\psi$ would be present in a full quantum mechanical
treatment, but as stated in the introduction that is beyond the scope of
 the present work. Finelli and Khlebnikov \cite{FK} also find such 
corrections in a different context where the fields are classical rather than in a 
 stochastic ensemble as here.},
$\Phi=\psi$ and we shall denote the metric
perturbation field by $\psi$. A friction mechanism will also be 
adopted to simulate the decay to the thermal gas of relativistic 
 particles dominating subsequent radiation era. These terms take the 
 simple form $\Gamma_{\phi}\phi'^{2}$ and $\Gamma_{\chi}\chi'^{2}$
and they can be consistently introduced into the equations of
motion in such a way that the Bianchi identities are 
 satisfied \cite{HM1}.

The stress-energy tensor and the equations of motion involve the 
averages of bilinear products of stochastic fields. Using as an 
example $\varphi_1$ these are defined as follows:
   \begin{equation}\label{eqIIA3}
 \langle\varphi_1(x,\tau)^2\rangle=(2\pi)^{-3}\int d^3k 
\varphi_{\bf k}\varphi_{\bf k}^*
\end{equation}
where $\varphi_{\bf k}$ are the k-mode functions entering in the 
expression
  \begin{equation}\label{eqIIA4}  \varphi_1({\bf x},\tau) = 
 \int \frac{d^{3}k}{(2\pi)^{\frac{3}{2}}} 
\lbrack e({\bf k})\varphi_{\bf k}(\tau)\exp(i{\bf k.x}) 
+  e^*({\bf k})\varphi_{\bf k}^*(\tau)\exp(-i{\bf k.x}) \rbrack, 
\end{equation}
with $e(\bf k)$ and $e^*(\bf k)$ being time-independent  
 $\delta$-correlated Gaussian variables \cite{PS} such that, where 
 $\langle .... \rangle$ denotes the average,
 \begin{equation}\label{eqIIA5}
 \langle e({\bf k})e^*({\bf k}') \rangle
 =\frac{1}{2} \delta^3({\bf k -k'});
 \end{equation}  
 with averages of $ee$ and $e^*e^*$ products being zero.The field
 $\psi({\bf x},\tau)$ is expressed as Eq.(\ref{eqIIA4}) with the 
 same gaussian variables but the mode functions being
 $\psi_{\bf k}(\tau)$. The stochastic gaussian variables of 
 $\chi$ are independent. Thus, for example,
 $\langle\varphi_1(x,\tau)\chi(x,\tau)\rangle$ is zero while
 $\langle\varphi_1(x,\tau)\psi(x,\tau)\rangle$ and
 $\langle\chi(x,\tau)^2\rangle$ are non-zero \cite{HM1}.
 
The thermal gas of relativistic particles is described by the 
hydrodynamical variables $\rho$, the density, and $p$, the 
pressure, related by the equation of state $p=\frac{1}{3}\rho$.
 As with the inflaton field the density has two components
\begin{equation}\label{eqIIA6}
 \rho({\bf x},\tau)=\rho_0(\tau)+\rho_1({\bf x},\tau),
 \end{equation}  
being the homogeneous background (classical) density and the
 non-homogeneous quantum or stochastic part. 
 As will be seen below the equations of motion imply that 
 the stochastic variables of $\rho_1$ are the same as those
of $\varphi_1$ and $\psi$. We assume that $\rho_0$ is zero 
at the beggining of reheating and only develops subsequently
through the friction mechanism $\Gamma_{\phi}\phi'^{2}$ and 
 $\Gamma_{\chi}\chi'^{2}$. In terms of consistency of the 
friction model the assumption that $\rho_0$ only becomes
 significant during the reheating is not unreasonable 
 because the friction only becomes significant when the fields
are oscillating rapidly.

            From the Lagrangian and from the hydrodynamical 
variables $p$ and $\rho$ we define the energy-momentum tensor
 \begin{equation}\label{eqIIA7}
T^{\mu}_{\nu} =\varphi^{,\mu}\varphi_{,\nu}+\chi^{,\mu}\chi_{,\nu}-
\lbrack\frac{1}{2}\varphi^{,\alpha}\varphi_{,\alpha}+
\frac{1}{2}\chi^{,\alpha}\chi_{,\alpha}-
V(\varphi)-V(\chi)-V_{int}-p\rbrack \delta^{\mu}_{\nu}
+ (\rho+p)u^{\mu}u_{\nu}
\end{equation}  
 where $u^{\mu}$ is the 4-velocity of the fluid.

 The homogeneous part of the total density, 
 $\rho_T({\bf x},\tau)$, is defined as 
$\rho_{0T}(\tau)=<T^0_0>$, giving, after ensemble averaging,
\begin{equation}\label{eqIIA8}
 \rho_{0T}(\tau) = \frac{1}{2 a^2} \lbrack \eta 
+a^2\bar{m}^2(\varphi_0^2+\langle \varphi_1^2 \rangle)
+a^2M^2 \langle \chi^2 \rangle
+\langle \varphi_{1,i}^2 \rangle+\langle \chi_{,i}^2 \rangle
 -4\varphi'_0\langle\varphi_1'\psi\rangle\rbrack +\rho_0,
\end{equation} 
\begin{equation}\label{eqIIA9}
 \eta =  
 \varphi_0'^2+\langle\varphi_1'^2\rangle+\langle \chi'^2 \rangle,
\end{equation} 
\begin{equation}\label{eqIIA10}
\bar{m}^2 \equiv m^2+g^2\langle \chi^2 \rangle.
\end{equation}             
The definition of $p_{0T}$ is slightly more subtle. Although we 
shall be using the Friedmann equation to calculate $a(\tau)$:
\begin{equation}\label{eqIIA11}
(a'/a)^2= \frac{8\pi G}{3}a^2\rho_{0T}(\tau).
\end{equation}
we should ensure that this equation is indeed compatible with the
second order (time-time) Einstein equation for $a''/a$ which is:
\begin{equation}\label{eqIIA12}
a''/a-(a'/a)^2= -\frac{4\pi G}{3}a^2(\rho_{0T}(\tau)+3p_{0T}(\tau)).
\end{equation}
Deriving Eq.(\ref{eqIIA11}) and comparing with Eq.(\ref{eqIIA12}) we find
 \begin{eqnarray}\label{eqIIA13}
 p_{0T}(\tau) & = & \frac{1}{2 a^2} \lbrack \eta_{-}
-a^2\bar{m}^2(\varphi_0^2+\langle \varphi_1^2 \rangle)
-a^2M^2 \langle \chi^2 \rangle
 -4\varphi'_0\langle\varphi_1'\psi\rangle\rbrack+\nonumber\\
\rho_0/3  & - & (a'/a)^{-1}\lbrack 2a^{-2}\varphi'_0\langle\varphi_1\psi'\rangle
 +\frac{4}{3}\langle\rho_1\psi'\rangle \rbrack. 
\end{eqnarray} 

\begin{equation}\label{eqIIA13b}
 \eta_{-}= \varphi_0'^2+\langle\varphi_1'^2\rangle+\langle \chi'^2 \rangle
-\langle \varphi_{1,i}^2 \rangle/3-\langle \chi_{,i}^2 \rangle/3.
\end{equation} 
 This differs, in the addition of the last term in square brackets, 
 from the expression obtained by using 
$p_{0T}(\tau)\delta^i_j = -\langle T^i_j \rangle$
 which we have used previously\cite{HM1}, but it has the additional virtue of
 precisely maintaining the continuity (or 'energy conservation') equation. 
 However in the equations of motion the expression for $p_{0T}$ only occurs
 in the metric perturbation equation (Eq.(\ref{eqIIA19}) below) and in fact
 the addition of the last term seems to make little difference
 to the numerical results.

Besides the Friedmann equation there are equations of motion for the other 
homogeneous  variables\footnote{Both equations can
be obtained from the Bianchi identities as shown in Appendix A 
 of ref.\cite{HM1}},  these being $\varphi_0$ and $\rho_0$:
 \begin{equation}\label{eqIIA14}
\varphi_0''+2(a'/a)\varphi_0'+a^2 \bar {m}^2 \varphi_0
-4\langle \psi'\varphi_1' \rangle-4\langle \psi\nabla^2\varphi_1\rangle
+2a^2\bar{m}^2 \langle \psi\varphi_1 \rangle
 =-a\Gamma_{\phi}(\varphi'_0+2\langle \psi\varphi'_1 \rangle)
\end{equation} 
\begin{equation}\label{eqIIA15}
\rho_0'+4(a'/a)\rho_0=
a^{-1}\Gamma_{\varphi}(\varphi_0'^2+\langle \varphi_1'^2 \rangle)+
a^{-1}\Gamma_{\chi}\langle \chi'^2 \rangle
-2a^{-2}\varphi_0'\langle \psi\nabla^2\varphi_1\rangle
+4\langle \psi'\rho_1\rangle
\end{equation}
There remain the spatially non-homogeneous equations which we write in 
the $k$-component form. These components are specified as the complex mode 
functions $\chi_k$, $\varphi_k$, $\psi_k$ and $\rho_{1k}$. Their wave number 
dependence, given  by the succeeding equations, is only on 
 $k \equiv \left|{\bf k}\right|$. There are 4 equations which we take as
 those for $\chi_k$, $\varphi_k$ and the time-time and space-space
 Einstein equations for $\psi_k$\footnote{As are $\varphi_k$ and $\psi$,
 $\chi$ is complex and the real and imaginary parts are not constant 
 multiples of each other for smaller occupation numbers of a mode}. 
(These being satisfied the time 
 derivative equation equation for $\rho_{1k} \equiv 3p_{1k}$ is implicit 
 through the Bianchi identities.)
\begin{equation}\label{eqIIA16}
\chi_k''+2(a'/a) \chi_k'+(k^2+a^2\bar{M}^2)\chi_k = 
 -a\Gamma_{\chi}\chi'_k 
\end{equation}
where $\bar{M}^2$ is a function of $\tau$ given by 
 \begin{equation}\label{eqIIA17}
\bar{M}^2(\tau)=
M^2+g^2\varphi_0(\tau)^2+g^2\langle\varphi_1({\bf x},\tau)^2\rangle,
\end{equation}

\begin{equation}\label{eqIIA18}
\varphi_k''+2(a'/a)\varphi_k'+(k^2+a^2 \bar {m}^2) \varphi_k -
4\varphi_0'\psi_{k}'+
2a^2\bar{m}^2\varphi_0\psi_k= 
 -a\Gamma_{\varphi}(\varphi'_k+2\varphi'_0\psi_k),
\end{equation}                  
 the equations actually holding for each separate value of
 ${\bf k}$; thus consistency of Eq.(\ref{eqIIA18}) justifies our previous 
 statements that $\psi_{\bf k}$, 
 the mode functions of the metric perturbation, should be associated with the 
 the same stochastic variables  (or quantum operators) as $\varphi_{\bf k}$.

The space-space Einstein equation for the metric perturbation can be
 written
\begin{equation}\label{eqIIA19}
\psi_k''+3\psi_k'(a'/a) +\psi_k(a'/a)^2= 4\pi G a^2 p_{Tk}
 +\frac{8\pi G}{3}a^2(\rho_{0T}+p_{0T})\psi_k
\end{equation}
 where $p_{Tk}$ is the $k$-component of the
non-homogeneous part of the total momentum:  

\begin{equation}\label{eqIIA20}
 p_{Tk} = a^{-2}\lbrack 
-\eta_{+}\psi_k 
+ \varphi_0'\varphi_k'
-2\varphi_k'\langle \psi\varphi_1' \rangle
 -a^2\bar{m}^2\varphi_0\varphi_k \rbrack
 +p_{1k},
\end{equation}

\begin{equation}\label{eqIIA20b}
 \eta_{+}= \varphi_0'^2+\langle\varphi_1'^2\rangle+\langle \chi'^2 \rangle
+\langle \varphi_{1,i}^2 \rangle/3+\langle \chi_{,i}^2 \rangle/3,
\end{equation}

The time-time Einstein equation is
\begin{equation}\label{eqIIA21}
3\psi_k'(a'/a) +\psi_k(k^2+3(a'/a)^2)= 4\pi G a^2 \rho_{Tk}.
\end{equation}

\begin{equation}\label{eqIIA22}
  \rho_{Tk} = a^{-2}\lbrack 
-(\eta-\langle\varphi_{1,i}\varphi_{1,i}\rangle
-\langle\chi_{,i}\chi_{,i}\rangle )\psi_k 
+ \varphi_0'\varphi_k'
-2\varphi_k'\langle \psi\varphi_1' \rangle
 +a^2\bar{m}^2\varphi_0\varphi_k \rbrack
 +\rho_{1k}.
\end{equation}

 Rather than using the mode equation for $\rho'_{1k}$ to find $\rho_{1k}$
we find it from Eqs.(\ref{eqIIA21},\ref{eqIIA22}) and then use the equation of state
$p_{1k}=\frac{1}{3}\rho_{1k}$ to solve the $\psi$ equation 
(\ref{eqIIA19},\ref{eqIIA20}).
\footnote{The $\rho_{1k}$ mode equation  is still maintained; the Bianchi
 identities hold and imply the $\rho_{1k}$ mode equation when the 
other equations of motion are explicitly fulfilled, as they are \cite{HM1}.} 
 
In addition there is the time-space Einstein equation which is not independent
of the other two Einstein equations but which acts as 
an equation of constraint on the initial values where we use it to fix the 
value of $\psi'$ at the beginning of reheating. It takes the form

\begin{equation}\label{eqIIA23}
 \psi'_k + (a'/a)\psi_k = 4\pi G\varphi_0' \varphi_k
 \end{equation}

\subsection{Initial conditions}
\label{subsecIIB}
We need to have the values of $H$, the fields and their 
 time-derivatives at the end of inflation to supply the initial conditions 
 for the reheating equations of motion . We have chosen to use a specific 
 inflationary model having the advantage that the solutions are 
 analytically expressible. This is power-law inflation with an exponential 
 potential: $V =  U\exp (-\lambda\varphi)$ where $U,\lambda$ are constants.
 This potential of the inflation era stands in place of the potential
 $V(\varphi)=\frac{1}{2}m^2\varphi^2$ of the
 reheating era but otherwise the Lagrangians are the same
 and in particular both have the interaction potential 
 $V_{int}(\varphi,\chi) = \frac{1}{2} g^2 \varphi^2 \chi^2$.
 We have ensured the correct continuity of the potential by imposing 
 the Lichnerowicz  conditions \cite{DER}, as well as using the equations of 
 motion and constraint appropriate to each era at the boundary \cite{HM1}.   

 That the $\chi$-field be  strongly suppressed in the inflation era,
 as referred to previously \cite{LLMW,BassV3,JS}, is  
 an essential feature in maintaining the analytic form of power-law 
 inflation. The many efold increase of $a$ during the inflationary era indicates 
 an exceedingly small value for $\chi_k$ at the beginning of reheat 
 \cite{LLMW,BassV3,JS}, which may be as little as $10^{-50}$ of its 
 initial value. This is an important qualitative feature of our
 initial conditions. We have taken the initial value of $\chi$ to be 
 $10^{-n}m_{Planck}^{-1/2}$ where $n$ is of the order of 30. 
 (For such small mode functions in the beginning of reheat the transition 
 to classical stochastic functions cannot yet be made and 
 the complex formalism, which is applicable also to the quantum 
 case \cite{PS,HM1} should be retained.

 For power-law inflation the scale factor $a \propto (\tau_i - \tau)^p$ 
 and we are free to choose $p$ by specifying $\lambda$ in the 
 potential \cite{HM1}. In the numerical results quoted we have chosen 
 $p=-1.1$ ($p=-1$ corresponds to exponential inflation).

Knowing the analytic solutions of the unperturbed equations enables the 
deduction of $\psi_k$ and $\varphi_k$ in the inflation era 
starting from the initial quantum oscillations\cite{HM1}. These are

\begin{equation}\label{eqIIB1}
\psi_k = a^{-1}\sqrt{\frac{4\pi G}{2k}}
\alpha\sqrt{\left|\gamma\right|}(\mu'_k-\alpha\mu_k)/k^2
\end{equation}
\begin{equation}\label{eqIIB2}
\varphi_k = a^{-1}(2k)^{-1/2}[\mu_k + \alpha\gamma (\mu'_k-\alpha\mu_k)/k^2]
\end{equation}
where $\gamma \equiv 1-\alpha'/\alpha^2=(p+1)/p$ and
\begin{equation}\label{eqIIB3}
\mu_k'' + (k^2-a''/a)\mu_k = 0
\end{equation}

Since $a''/a=\frac{p+1}{p} (\tau_i - \tau)^{-2}$, Eq.(\ref{eqIIB3}) can be 
expressed as a Bessels equation in terms of
\begin{equation}\label{eqIIB4}
y \equiv  k(\tau_i-\tau)=\frac{k}{aH}\left|p\right|
\end{equation}
and the solutions which satisfy the asymptotic conditions for a scalar field
 are
\begin{equation}\label{eqIIB5}
\mu_{k}(y) = \sqrt{\frac{\pi y}{2}}(J_{\nu}(y) - iY_{\nu}(y))
\exp\lbrack-i(\frac{1}{2}\nu\pi + \frac{1}{4}\pi)\rbrack
\end{equation}
where $\nu=1/2-p$.

For $k/a \ll H$ (many times over fulfilled by the $k$ relevant  
to the CMBRF observations)
$\mu_{k}(y) \propto y^p \propto k^p$ and if this $k$-dependence were to pass 
unaltered through the reheat era into the radiation era it would yield a 
power-law spectrum scaling as $k^{n-1}$ where $n$, the spectral index, is given
 by $n=2p+3$.

  From the conditions of continuity  it follows that at the junction of 
inflation and  reheat  
 \begin{equation}\label{eqIIB6}
     H = \sqrt{4\pi G}m\varphi_0 \sqrt{p/(2p-1)}       
\end{equation}
 and this, using Eq.(\ref{eqIIB4}), determines the value of $y$ at the 
 junction 

 \section{SOLVING THE REHEATING EQUATIONS}
 \label{secIII}
            One of the main aims of the present work is 
 to see whether,and if so how far, parametric
 resonance of a second scalar influences the fluctuations of the CMBR.

            An important parameter is $\zeta_k$ \cite{BST,LYTH} which,
 for $k^2/a^2 \ll H^2$, is equivalent to the curvature perturbation
 ${\cal R}_{k}$. $\zeta$ is defined in terms of the gauge-invariant metric
 perturbation $\Phi$, equal in our case to the longitudinal gauge
 perturbation $\psi$; in terms of this variable we have \cite{MUK}
\begin{equation}\label{eqIII1}
\zeta_k \equiv \frac{2}{3}(H^{-1} \dot\psi_k + \psi_k)/(1+w) + \psi_k  
 \approx -{\cal R}_{k} 
\end{equation}
 where $w=p_{0T}/\rho_{0T}$, the total (homogeneous) pressure to density ratio.
 If $k^2/a^2 \ll H^2$ then $\zeta$ is constant during the 
 radiation era and constant if there be only one scalar field, the inflaton, during
 inflation and also during reheating if the inflaton perturbation can be treated 
 to first order. That is, as it is often expressed, it is constant
 from horizon exit during inflation to horizon re-entry around the beginning
 of the matter era.The question arises whether, with another scalar
 field having parametric resonance during reheating, can $\zeta_k$ exit 
 reheating with a different value from that with which it enters  
 \cite{LLMW,JGBDW,WMLL}  As we shall see, for values of $g/m$ bigger than
 about $5000$ there is parametric resonance, with $\chi$ comparable to the  
 other scalar densities, and $\zeta$ is 
 larger than its initial value\footnote{This is a correction to some
 numerical values given in our previous work. Due to a misprint in the
 computing code some decreasing values of $\zeta$ for large couplings
 were, wrongly, found.}  and also there can be contributions to this effect
 from higher order inflaton perturbations.
 Then, knowing $\zeta$ at the end of reheating (that is at the beginning
 of the radiation era, asessed as in the succeeding paragraph) we can use 
 known methods to find some properties of the fluctuations and assess 
 consequences of the change in $\zeta$ in reheat.

 Another method \cite{HMM,GRI} is to trace 
 the metric perturbation $\psi_k$, for the 
 appropriate very small values of k, from the end of reheating to the matter
 era where the Sachs-Wolf effect comes into play.  Once the system 
 thermalizes, with the gas of 
 relativistic particles obeying an equation of state of the form 
 $p=\rho/3$, we can check that as the
 reheating era tends to the beginning of the radiation era then  
 $\psi_k=b1 + b2/a^3, a(\tau) \approx \tau$,
 is a solution of Eqs. (\ref{eqIIA19}),(\ref{eqIIA21}), for $k^2 \approx 0$,
  where the constant 
 b1 has to be determined from the computations. The attainment of constant value 
 for $\psi_{k\approx 0}$ and $\rho a^4$ will be taken as indicating the 
 beginning of the radiation era;  $\psi$ can then be traced throughout the 
 radiation era from its analytic form which holds therein. 

In this paper we evaluate the power spectrum of the fluctuations and find 
 the quadrupole moment (which can give the COBE normalization) using the 
first method above. 

 In contrast to our results given below, the work of refs.\cite{LLMW} finds
 very small effects of $\zeta$ variation by expressing $\dot\zeta$ in terms 
of the non-adiabatic pressure variation. This comes using the reasonable 
physical assumption - based on the approximate validity of the 
Robertson-Walker metric - that after smoothing on a cosmological 
scale below the ones of interest the spatially inhomogeneous 
variables such as $\rho_k, p_k$ can be treated as 
perturbations. There is no evident agreement between that 
expression for $\dot\zeta$ and Eq.(\ref{eqIIIA1}) below which uses no 
assumptions on $\rho_k, p_k$ being perturbative. 
However, as we discuss in sections I and V, in our work 
we do consider the variation $\psi$ from the Robertson-Walker metric
to be perturbative.

  \subsection{Small coupling constant, g/m}
 \label{subsecIIIA}
 
            Having derived in Sec. \ref{subsecIIB} the expressions allowing the
 magnitude of the relevant modes at the beginning of reheat to be 
calculated, thus fixing the input of $\psi, \varphi_1$ and $\chi$ to the reheat 
era, we may proceed with the numerical integration of the reheating
equations of Sec.\ref{subsecIIA}, taking the integration up to a point where we 
may consider the system to be well thermalized. The output of this 
numerical development will then be used in Sec.\ref{secIV}.

            Unless otherwise stated, the results are for $p=-1.1$,
 $M/m=0.02$, $\phi_0=.3m_{Planck}$ at the beginning of reheat 
 and the inflaton mass   $m=10^{-7}m_{Planck}$ ($m_{Planck}=G^{-1/2}$).
 The values for the coupling constant $g$, in 
 $\frac{1}{2}g^2\varphi^2\chi^2$, will 
 be quoted in terms of $g/m$. 

 The frictional decay constants $\Gamma_{\chi}$, $\Gamma_{\varphi}$  
 have been investigated in the range $0.1>\Gamma_{\chi}/m>0$ and 
  $0.005>\Gamma_{\varphi}/m>0$, larger values tending to suppress the 
 resonance. However the results presented here will be for small values
 of the order of $10^{-4}$. The reason is that then the parametric 
 resonance has had time to reach its full effect before the 
 $\varphi_0, \varphi_1$ and $\chi$ decay. For larger decay constants
 the transition to the radiation era occurs when $\zeta$ is still 
 fluctuating very strongly leading to more  random $\zeta$ values in the
 radiation era. It is significant that 
 without this physical assumption the theory would be 
 very randomly dependent on the precise magnitude of the coupling.

            In this subsection we consider values of $g/m<4000$ and some 
 typical results are given in Table I.The main  
conclusion seems to be the fact that the end of reheating 
\footnote{The suffices 
$in$ and $out$ denote the beginning and end of reheat respectively}
values of our 
variables $\psi_{k \approx 0}$ and $\zeta_{k \approx 0}$ are practically 
 independent of $g/m$ (though not of the $\Gamma$'s which govern the rate
 of thermalization). This was checked varying $g/m$ including 
$g/m=0$, obviously a  no-resonance case, up 
to nearly $g/m=4000$. Such results, at first surprising, may be understood
looking carefully at the energies associated with the fields 
$\varphi_0, \varphi_1$and $\chi$. 

 The graphs in Figs, 1,2 and 3 show the reheating transition for $g/m=3000$
 and $\Gamma_{\chi} = \Gamma_{\varphi} = 5\times10^{-5}$.
 Fig.1 shows the energy densities of the inflaton particles $\varphi_1$ 
 and of $\chi$, displaying the parametric resonance. The scale is such
 that in all the graphs not all the rapid oscillations are adequately displayed.
 Fig. 2 shows the considerable $\zeta$ oscillations during the transition 
 and that it is rather constant in the initial period when, as we can see 
 from Fig. 3, the density of $\varphi_1$ is small compared to $\varphi_0$.
 Despite considerable oscillations to larger $\zeta$ the final result for 
 these small decay constants is that $\zeta$ decreases and by relatively little.
   
 Although $<\chi^2>$ increases by many powers of ten during
the resonance (more than $10^{20}$, for $g/m=3000,m=10^{-7}$, 
the energy associated with this field is, nevertheless,
always smaller than the energy of the fields $\varphi_0, \varphi_1$, which 
 thus control the dynamics of the system.
 The most striking feature is that the case $g=0$ having essentially zero
 $\chi$ field produces the same change in $\zeta$. This has already
 been found and discussed in ref.\cite{HM2} for the case where there is no
 particle decay ($\Gamma_{\chi}= \Gamma_{\varphi} = 0$). There the analytic 
 expression for the rate of change $\dot\zeta$ is non-zero due to the binary 
 (non-perturbative) terms in the inflaton particle field, $\varphi_1$ even
 if the $\chi$ field were zero:
\begin{equation}\label{eqIIIA1}
\frac{3}{2}H(1+w)\dot \zeta=\ddot{\psi}+H\dot \psi+2\dot H \psi-
(\dot \psi+H\psi)\frac{d}{dt}\ln(\rho_h+p_h),
\end{equation}

\begin{equation}\label{eqIIIA2}
\frac{d}{dt}\ln(\rho_h+p_h) =
2\frac{\dot{\varphi}_0\ddot{\varphi}_0+
\langle\dot{\varphi}_1\ddot{\varphi}_1\rangle+
\langle\dot{\chi}\ddot{\chi}\rangle
+\langle\dot{\varphi}_{1,i}\ddot{\varphi}_{1,i}\rangle/3+
\langle\dot{\chi}_{,i}\ddot{\chi}_{,i}\rangle/3 }
{\dot{\varphi}_0^2+\langle\dot{\varphi}_1^2\rangle+\langle\dot{\chi}^2\rangle
+\langle\dot{\varphi}_{1,i}\dot{\varphi}_{1,i}\rangle/3+
\langle\dot{\chi}_{,i}\dot{\chi}_{,i}\rangle/3 },
\end{equation}
 Thus the presence of the inflaton particles, comparable in their contribution 
 to the energy to the residual homogeneous inflaton field, generates a 
 change in $\zeta$ in the reheating - or more exactly the preheating - period.    
 However as we see from Table I, for the small value of $\Gamma_{\varphi}$
 chosen (for the reasons given above), the ultimate change in $\zeta$ is
 quite small.

 \subsection{Large coupling constant, g/m}
 \label{subsecIIIB}

  We see in Table II  that  $\zeta$ increases in this regime:
 $\zeta_{out}>\zeta_{in}$.
 The graphs in Figs, 4, 5 and 6 show the reheating transition for $g/m=4700$
 and $\Gamma_{\chi} = \Gamma_{\varphi} = 5\times10^{-5}$.
 We may note that: firstly  the parametric resonance 
 in $\chi$ is greater as one would expect with a larger coupling;
 secondly that $\zeta_{out}$ now varies with $g$. Thirdly the
 fluctuations in $\zeta$ tend to be larger, with a considerable increase 
 in a  smoothed value of $\zeta$ before true heating takes over, the increase
 taking off when the $\chi$ density attains the $\varphi_1$ density  as 
 can be seen from Figs.4 and 5. Again we should emphasize the role of the 
 decay widths; if these had been much larger in that case then 
 the final $\zeta$ would have been much changed.

 \subsection{The sudden approximation}
 \label{subsecIIIC}    

It may be of interest to compare our results with those of the sudden 
approximation, where there is no use of the reheating mechanism, 
especially because of the independence of $g$ in a range of smaller 
couplings set out in section \ref{subsecIIIA}.  
The matching conditions, deduced by Deruelle and Mukhanov
 \cite{DER} from the Lichnerowicz conditions, imply \cite{HMM} 
that $\psi$ and the function
\begin{equation}\label{eqIIIC1}
 \Gamma \equiv (\psi' \alpha^{-1} + \psi 
 -\frac{1}{3}\nabla^2 \psi \alpha^{-2})/(1-\alpha' \alpha^{-2}), 
\end{equation}
 as well as $\alpha=a'/a$,
be continuous on the sudden transition hypersurface, say $\tau=\tau_2$. 
 At the end of inflation, that is on one side, denoted $\tau_{2-}$, of the 
transition  hypersurface, the values of $\psi$ and $\Gamma$ may be
 written  
\begin{equation}\label{eqIIIC2}
 \psi(\tau_{2-}) = \frac{p+1}{2p+1}\Sigma, 
\end{equation}
\begin{equation}\label{eqIIIC3}
 \Gamma(\tau_{2-}) = \frac{p}{2p+1}\Sigma, 
\end{equation}
where $\Sigma \propto y^p$ is given by the inflation model as in
 section \ref{subsecIIB}. On the other side of the hypersurface
 in the radiation era, we may write the general solution for $\psi$
 \begin{equation}\label{eqIIIC4}
\psi_k =  [B_1(\epsilon\cos\tilde{\epsilon}-\sin\tilde{\epsilon}) + 
 B_2(\cos\tilde{\epsilon} + \epsilon\sin\tilde{\epsilon})]/a\epsilon^2,
\end{equation}  
where $\tilde{\epsilon} =\epsilon - \epsilon_2$, $\epsilon = k\tau/\sqrt{3}$ 
 and $\epsilon_2=k\tau_2/\sqrt{3}$ 
 is the value of $\epsilon$ at the beginning of the radiation era.(We take 
 the convention on $\tau$ that $a=c\tau$, $c$ constant, in the radiation era.)
 From Eq.(\ref{eqIIIC4}) we can obtain the values of $\psi$ and $\Gamma$
 on the radiation era side of the hypersurface denoted $\tau_{2+}$. Then 
 solving  
 \begin{equation}\label{eqIIIC5}
 \psi(\tau_{2+}) = \psi(\tau_{2-}) , 
\end{equation}
\begin{equation}\label{eqIIIC6}
 \Gamma(\tau_{2+}) = \Gamma(\tau_{2-}), 
\end{equation}
gives
\begin{equation}\label{eqIIIC7}
 B_1\epsilon_2 = -2a_2 \Sigma + 2\epsilon_2^2 a_2 \frac{p+1}{2p+1}\Sigma, 
\end{equation}
\begin{equation}\label{eqIIIC8}
 B_2 = 2a_2 \Sigma - \epsilon_2^2 a_2 \frac{p+1}{2p+1}\Sigma
\end{equation}
 For values of $k$ relevant to the CMBRF we can expand $\psi$ in powers of 
$\epsilon$ through the pure radiation era:
\begin{equation}\label{eqIIIC9}
 \psi(\tau) = \psi_{qa} + O(\epsilon_2^2/\epsilon^2)
\end{equation}
\begin{equation}\label{eqIIIC10}
 \psi_{qa} = \frac{2}{3}\frac{\epsilon}{\epsilon_2}\frac{a_2}{a}\Sigma
           =\frac{2}{3}\Sigma
\end{equation}
where in the last equation we have used the fact that both $a$ and 
$\epsilon$ are proportional to $\tau$. Thus, in the radiation era,
$\psi$ changes rapidly at the beginning and after a few efolds of $a$
takes up a constant quasi-asymptotic value, $\psi_{qa}$. This constancy 
of $\psi$ in most of the radiation era is a general property and 
not peculiar to its initiation by a sudden transition. For the sudden 
transition one finds that $\psi_{qa}$ is considerably bigger than 
$\psi_{in}$, the value at the end of inflation:
\begin{equation}\label{eqIIIC11}
 \psi_{qa} = \frac{2}{3}\frac{2p+1}{p+1}\psi_{in}
           =7.33\psi_{in}.
\end{equation}
 The corresponding quasi-asymptotic value found in our gradual reheating 
calculations for $g/m < 4500$ is surprisingly close:
\begin{equation}\label{eqIIIC12}
 \psi_{qa} \approx 7.0 \psi_{in}.
\end{equation}

A  similar explicit evaluation of $\zeta$ in the 
sudden transition case gives from the beginning of  and all through the
radiation era   
\begin{equation}\label{eqIIIC13}
 \zeta  = \frac{\epsilon}{\epsilon_2}\frac{a_2}{a}\Sigma =\Sigma
        = \zeta_{in}.
\end{equation}
 In the gradual reheating calculations for $g/m < 4500$

\begin{equation}\label{eqIIIC14}
 \zeta  = \zeta_{out} = .88 \zeta_{in}.
\end{equation}
Thus when parametric resonance is small our gradual transition calculation
 and the sudden transition result approximate each other to within about 12\%
 for $\zeta$ and 5\% for $\psi$.

 \section{CMBR FLUCTUATIONS; QUADRUPOLE MOMENT}
 \label{secIV}
We consider that the $\zeta$ effects we have calculated for a non-linear 
 theory would apply, qualitatively,  to a large range of parameters. Nevertheless
we should show that the parameters used here are quite near to ones 
which correctly normalise the theory to the cosmic microwave background 
 fluctuations as required for modelling the real world.

            So in this section we shall be concerned with the calculation 
of the scalar perturbations in the CMBR, through the consideration of 
the Sachs-Wolf effect, describing the fractional variation of the 
temperature $dT/T$. Our procedure will not involve details of the radiation 
to matter era transition or of the matter era itself and so will only be
 applicable to the lower multipoles of the radiation; in particular
we shall calculate the quadrupole moment to obtain the CMBR normalization.
 Here we briefly describe how we implement a rather standard method
(see for example ref.\cite{L&L}) using the power function of $\psi$.
(We have also calculated using another method, that of ref.\cite{HMM},
 which takes account of some details and have obtained similar results.) 
We track the development of the metric perturbation $\psi$, from the end of 
reheating, through the radiation era and into the matter era. 

The cosmic microwave background is described by a single number, the 
 temperature $T$ of the black body radiation. The 
 small anisotropies in it from the direction dependent 
 fluctuations, $\delta T({\bf e})$,  where ${\bf e}$ is a unit
 vector in the direction of observation, depend on the metric 
 perturbation along the light path after emission at last scattering
 - the Sachs-Wolfe effect. This can be expressed as \cite{SW,L&L}
\begin{equation}\label{eqIV1}
\frac{\delta T}{T}({\bf e}) = -\frac{1}{5}{\cal R}({\bf x}_{LS},\tau_{LS}) 
\end{equation}
 where ${\bf x}_{LS}=\frac{2}{H_0}{\bf e}$, $\frac{2}{H_0}$ being a good 
 approximation to the distance of the last scattering sphere with 
 $H_0$ the present Hubble and $\tau_{LS}$ the emission time.
 ${\cal R}({\bf x}_{LS},\tau_{LS})$ is stochastic being given linearly by   
 $\psi({\bf x}_{LS},\tau_{LS})$ which is itself stochastic being 
 expressed by the corresponding formula to Eq.(\ref{eqIIA4}).

 Using Eq.(\ref{eqIV1}) the angular correlation moments, ${\cal C}_l$, 
 can be calculated 
 in terms of ${\cal R}_k(\tau_{LS})$. With ensemble averaging using 
 Eq.(\ref{eqIIA5}) it can be shown, after some manipulation \cite{L&L}, that
 \begin{equation}\label{eqIV4} 
 {\cal C}_l=\frac{4\pi}{25}\int\frac{dk}{k}{\cal P}_{{\cal R}}(k)
\left[ j_l\left(\frac{2k}{a_0 H_0}\right)\right]^2
\end{equation} 
where ${\cal P}_{{\cal R}}(k)$ is the power function of the curvature perturbation:
\begin{equation}\label{eqIV5} 
 {\cal P}_{}(k)=\frac{k^3}{2\pi^2}|{\cal R}_{k}|^2
\end{equation} 

To calculate low moments we only need ${\cal R}_k(\tau_{LS})$ for values of
 the order of $a_0H_0$ and at that time they are well outside the horizon,
 $k \ll aH(\tau_{LS})$ and so ${\cal R}_k \approx -\zeta_k$ to a good approximation.
 So we can replace ${\cal R}$ by $\zeta$ in Eqs.(\ref{eqIV4}) and
 (\ref{eqIV5}.  
 Knowing $\zeta_k$ at the end of inflation for an appropriate range of $k$ 
 enables the calculation of correlation moments using these equations. 
This can be done in principle by computing
 $\zeta_k$ for sufficient $k$-values to cover the relevant range defined by the 
 spherical Bessel functions. Alternatively we note from section \ref{subsecIIB} that
 $\psi_k,\zeta_k \propto k^{p-1/2}$ at the beginning of reheating. The values of 
 $k/aH$ in the reheating period are extremely small 
 for values of $k$ appropriate to the CMBRF (and indeed even much larger values)
 and so the
 $k^2$ terms in the reheating equations are negligable. Thus the $k$-dependence  
 passes unaltered through the reheating  and we use it to perform the 
 $k$-integration  of Eq.(\ref{eqIV4}) to produce the results which now follow.

 We calculate the quadrupole moment defined, with $l=2$, as
\begin{equation}\label{eqIV9} 
 Q = T_0\sqrt{\frac{2l+1}{4\pi}{\cal C}_l}. 
\end{equation}   
 Let $R_{\zeta} \equiv \zeta_{k,out}/\zeta_{k,in}$, this ratio being independent
 of $k$ for the relevant $k$ and is found from the numerical solution of the reheat 
 equations (Tables I and II). In terms of the $k$-independent quantity
 $R_{\zeta}\zeta_{k,in}k^{1/2-p}$ we find
\begin{equation}\label{eqIV10} 
 {\cal C}_2=I_p\left[R_{\zeta}|\zeta_{k,in}|k^{1/2-p}
 \left(a_0H_0 \right)^{p+1} \right]^2
\end{equation}   
\begin{equation}\label{eqIV11} 
 I_p=\frac{2}{25\pi}\int\frac{dz}{z}z^{2p+2}j_2(2z).
\end{equation}    
 From Eqs. \ref{eqIIB1}-\ref{eqIIB6}

\begin{equation}\label{eqIV12} 
 |\zeta_{k,in}|k^{1/2-p}\left(a_0H_0 \right)^{p+1}=
 \sqrt\frac{2\pi p}{p+1}\left(\frac{a_0H_0}{a_{in}H_{in}}\right)^{p+1}
 \frac{H_{in}}{m_{Planck}}\vert \tilde M(p)\vert
\end{equation}     
\begin{equation}\label{eqIV13} 
\tilde M(p)=\frac{\sqrt\pi (-p/2)^p}{\sin((1/2-p)\pi)\Gamma(p+1/2)} .
\end{equation}    
 For $g/m < 4500$ and for the values of the decay constants 
 given in Table I we find  
\begin{equation}\label{eqIV14} 
 Q \approx 12\mu K. 
\end{equation}   
If $\zeta$ had passed unchanged through reheating we would have found
\begin{equation}\label{eqIV15} 
 Q \approx 14\mu K 
\end{equation}   

 The observed value is $ Q \approx 18 \mu K$, and our theoretical result can 
readily be changed to agree with this by adjustment of the parameters
 $\varphi_{0,in}$  - the homogeneous inflaton field at the end of inflation - 
 and $m$, the reheating potential parameter. Similar relatively small 
adjustments to these parameters (while scaling $g$,$M$,
$\Gamma_{\chi}$ and $\Gamma_{\varphi}$  proportionately to $m$)
can give agreement with the observed $Q$ for larger couplings such as
 those of Table II.

  Of course, while the evaluation above verifies that the parameters we 
 have used are in the right region, the point 
 is not the precise values of $Q$ found but rather the amount of change induced 
 in $\zeta$ by the reheating which forces changes in the parameters of the theory.

 \section{SUMMARY AND DISCUSSION}
 \label{secV}

 We have used  a well-known simple theory that can give rise to parametric resonance 
 of a scalar field, $\chi$, coupled to the scalar inflaton, $\varphi$, through a term 
 $g^2\varphi^2\chi^2$.  We have investigated preheating, followed by reheating 
 mediated by decay terms $\Gamma_{\varphi}\dot\varphi^2$ and 
 $\Gamma_{\chi}\dot\chi^2$  in the Lagrangian which implement a smooth
 transition to a radiation era with a hydrodynamic fluid of relativistic particles.
 We have arranged a transition from an inflationary era ro the heating era that 
 is also smooth. Besides non-linear effects of the extra field $\chi$ we have included
non-linear effects of the inflaton particle field $\varphi_1$. The investigations centre
 on the three-curvature related parameter $\zeta_k$ for values of $k$ relevant to 
 the cosmic microwave background fluctuations. While, for such $k$,  $\zeta_k$ is constant through
 the latter part of inflation and also through the radiation and subsequent eras as far
  as the generation of the fluctuations,  the results presented show $\zeta_k$ varying 
  in the heating transition. Thus they emerge into into the radiation era with a value 
  changed from that in inflation which implies a change in the fluctuations - or rather 
 a change in the parameters of the theory  from what they would have been if $\zeta$ 
 had been constant throughout. The change depends markedly on the magnitudes 
 of $g^2$,  $\Gamma_{\varphi}$ and $\Gamma_{\chi}$. 

 For given values of $\Gamma_{\varphi}$ and $\Gamma_{\chi}$ there is a range of 
 $g^2$ from small up to some value depending on the the magnitudes of 
$\Gamma_{\varphi}$ and $\Gamma_{\chi}$ where parametric resonance occurs
 but where the change in $\zeta$ is always by the same proportion. Though in the results
 presented this change is a diminution only of the order of 10\% or so, it is interesting 
 that it appears not to be due to the parametric resonance in $\chi$ but only to non-linear 
 effects of the inflaton particles produced - the energy density of the $\chi$ particles
  never quite reaching that of $\varphi$ particles. This change in $\zeta$ 
 should be no great surprise as when $<\varphi_1^2>$ is comparable with $|\varphi_0^2|$, 
 as it is during the latter stages of heating,
 theorems on the constancy of $\zeta$ break down. However there is a range of $g^2$,  
 $\Gamma_{\varphi}$ and $\Gamma_{\chi}$ where there is considerable parametric 
 resonance but the view that $\zeta$ should be unvarying throughout is a reasonable
 approximation.

 With  larger $g^2$ then $\zeta$ increases by a considerable proportion this being up to a 
 factor of 20 in the examples we have given.  These cases are characterised by the $\chi$
 energy density, after the onset of parametric resonance, approaching and at times 
surpassing the $\varphi_1$ energy density. As can be seen from the figures , $\zeta$ 
 fluctuates markedly after 2 or 3 efolds into the heating transition. If we make the $\Gamma$s
 larger so that the decay to radiation is quicker  then the variability of the radiation era 
 $\zeta$ with $g^2$ increases.

 In the inflationary period the $\chi$ field is restricted to within about the primordial value 
 of quantum fluctuations. As seen from Figures 1 and 3 this persists for a few efolds into the 
 heating transition before parametric resonance sets in and raises the energy density by about 
 50 powers of ten in half an efold. Subsequently the $\chi$ energy density never greatly
 exceeds that of the $\varphi_1$; it seems that the direct interaction between these two 
 fields is a strongly limiting factor. That is, the same interaction that gives rise to parametric 
 resonance may also act to limit it because it contains inflaton particles as well as the 
 classical inflaton field, $\varphi_0$. This feature seems likely to apply to many other 
  models with parametric resonance. So then the inflaton particles always play an important 
 role in the evolution of $\zeta$ and the revelation of their significance may be the most 
 important result of the non-linear theory.     

 The formalism we use has been explained at much greater length in a preceeding 
 paper \cite{HM1}. In that paper also fairly strong $\zeta$ variations were computed
 but the results given here  supercede (see Section \ref{secIII}) 
 those previous ones \cite{HM1,HM2}. Also the present work extends the previous 
 work by treating  the completion of reheating into the radiation era where $\zeta$ 
 becomes a constant, different from that at the end of inflation, and thus the consequences 
 for the cosmic microwave background fluctuations become evident. Our model extends 
 smoothly from the primordial quantum excitations to the creation of the fluctuations.

 The main feature of our formalism is that as the number density of the original quantum 
 fields increases a smooth transition to where they can be treated as classical stochastic 
 occurs \cite{PS}; the only manageable treatment of the non-linear terms in the equations
 of motion is to take ensemble averages. This gives rise to loop imtegral factors, such as
 that of Eq.(\ref{eqIIA3}), in the equations of motion. The evaluation of these involves 
 both an infra-red, low  $k$, cut-off and an ultra-violet, high $k$, cut-off. This has been 
 discussed in detail in Appendix B of ref.\cite{HM1} where the low $k$ cut-off was 
 regarded as a contribution to the renormalization of the homogeneous part of the field.
 The high $k$ cut-off is fixed to correspond to a wavelenth $H^{-1}$, the inverse Hubble
 length at the beginning of heating. Such a cut-off has been used in the work of other 
 authors for example in ref.\cite{LLMW}. Though the only obvious approximate 
 magnitude it is a major source of uncertainty in our work, as any cut-off very much less
 or very much greater would significantly affect our results. A very much smaller cut-off
 of course corrsponds cloosely to the neglect of all non-linear effects.   

 Where, on the contrary,  we have used a linear approximation is in working in the 
 equations of motion to first order in the metric perturbation $\psi$. The measure of 
 validity we have taken is, in accord with the metric Eq.(\ref{eqI1}), to compare
 $2\sqrt{<\psi^2>}$ with unity\cite{HM1}. In cases where $\zeta$ decreases this 
corresponds throughout the heating process to a less than 1\% perturbation to 
 the metric and the first order perturbation theory is well justified. But in cases 
 where $\zeta$ increases it can reach up to nearly a 10\% perturbation in the latter 
 stages of reheating. While we do not think that this gives doubt that these are cases 
 where $\zeta$ increases markedly nevertheless the quantitative aspect would benefit from 
 a more refined treatment. 

 We find it difficult to see any direct 
 observational test of the changes in the heating transition.
 In the body of the paper we have used that the shape of the power spectrum, in our theory
  a power law, emerges unchanged from the transition and so only depends on the 
 inflationary theory. The parameters of nearly all envisaged theories can be changed to 
 follow the magnification (or diminution) of $\zeta$ and adjust to the COBE or other 
 normalization. The dependence on the details of preheating and reheating makes this 
 adjustment uncertain in the present state of knowledge. The best chance may lie in 
 a comparative calculation of gravitational wave production in the hope that this can
 be observed \cite{KT2,BAB}.

\begin{table}
\caption{The change in $\zeta$ in the inflation to radiation era 
 transition, shown for smaller values of the $\chi,\varphi$ coupling, $g$.
 Also shown are the increase, $n(efolds)$, in $\ln(a)$ in the transition
 and the ratio of densities of $\chi$ and $\varphi_1$ at the
 maximum density of $\chi$.  The decay constants are given by
$\Gamma_{\chi}/m=\Gamma_{\varphi}/m=5\times10^{-5}$.}  
\label{TI} 
\begin{tabular}{|c|c|c|c|} 
$g/m$  & $n(efolds)$ & $ \log(den\chi/den\varphi_1) $ 
& $ \zeta_{out}/\zeta_{in} $\\ \hline
$1000$ & $7.9$ & $-39$ & $0.88$\\ 
$3000$ & $7.6$ & $-3.9$ & $0.87$\\
$4000$ & $7.9$ & $-14$ & $0.88$\\
\end{tabular}
\end{table}

\begin{table}
\caption{The increase of $\zeta$ in the inflation to radiation era 
 transition, shown for some larger values of the $\chi,\varphi$ coupling, $g$.
 Also shown are the increase, $n(efolds)$, in $\ln(a)$ in the transition
 and the ratio of densities of $\chi$ and $\varphi_1$ at the
 maximum density of $\chi$.  The decay constants are given by
$\Gamma_{\chi}/m=\Gamma_{\varphi}/m=5\times10^{-5}$.}  
\label{TII} 
\begin{tabular}{|c|c|c|c|} 
$g/m$  & $n(efolds)$ & $ \log(den\chi/den\varphi_1) $ 
& $ \zeta_{out}/\zeta_{in} $\\ \hline
$4700$ & $7.9$ & $-.9$ & $5.6$\\ 
$4900$ & $7.9$ & $-.9$ & $10.2$\\
$5000$ & $7.6$ & $-.5$ & $19.1$\\
$7000$ & $7.5$ & $+.7$ & $6.3$\\
$12000$ & $7.9$ & $-.9$ & $10.3$\\
\end{tabular}
\end{table}

\begin{figure}
\begin{center}
\input{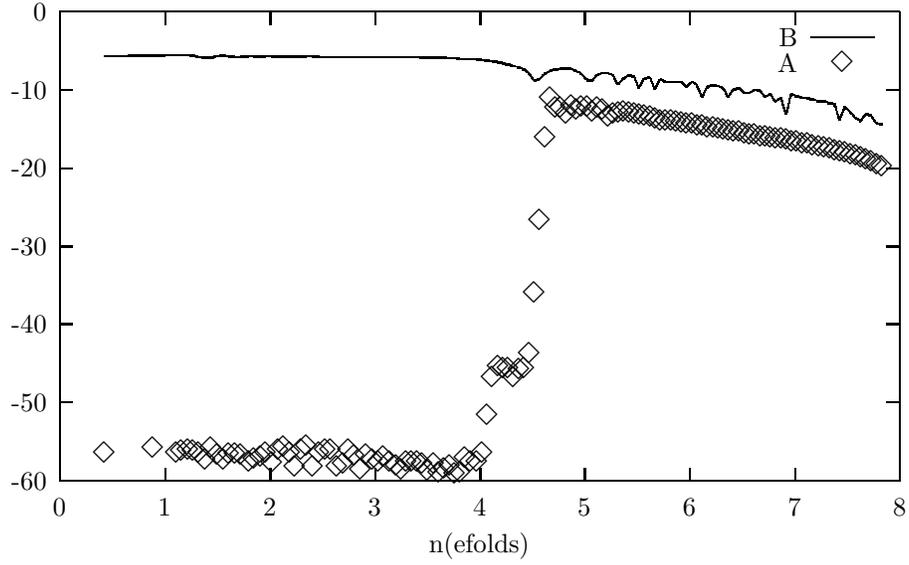}
\end{center}
\caption{For $g/m=3000$,
 logarithmic plot of energy densities of $\chi$ (A)and
 $\varphi_1$ (B) versus the number of e-folds of expansion 
 in the reheat era. In all figures densities are in units
 $m^2\times m_{Planck}^2$, where $m=10^{-7}m_{Planck}$.}   
\end{figure}

\begin{figure}
\begin{center}
\input{ABHfig2.tex}
\end{center}
\caption{For $g/m=3000$,
 plot of $\lg(\zeta/\zeta_{in})$ versus the number of e-folds of
 expansion in the reheat era.}   
\end{figure}

\begin{figure}
\begin{center}
\input{ABHfig3.tex}
\end{center}
\caption{For $g/m=3000$,
 logarithmic plot of energy densities of 
 $\varphi_1$ (A), $\varphi_0$ (B) and $\rho_0$ 
 versus the number of e-folds of expansion 
 in the reheat era.}   
\end{figure}

\begin{figure}
\begin{center}
\input{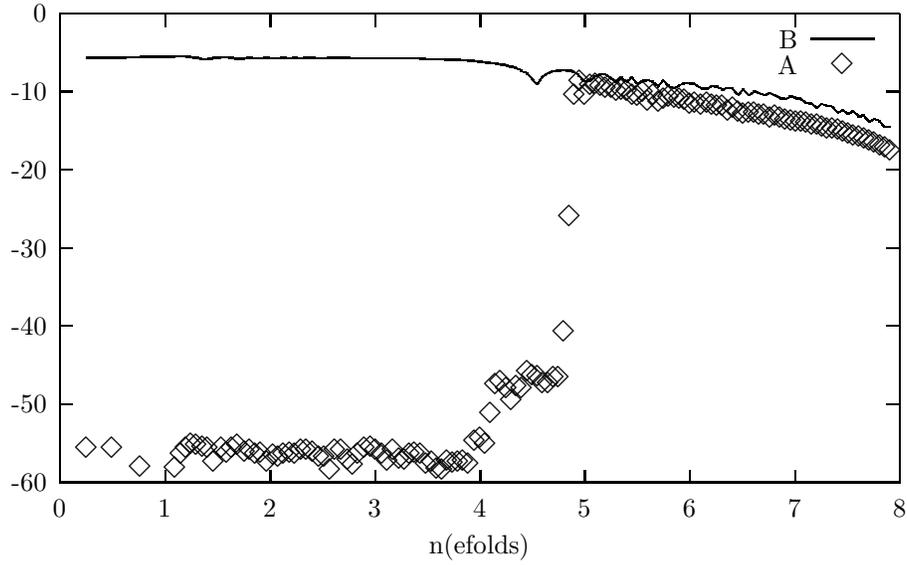}
\end{center}
\caption{For $g/m=4700$,
 logarithmic plot of energy densities of $\chi$ (A)and
 $\varphi_1$ (B) versus the number of e-folds of expansion 
 in the reheat era.}   
\end{figure}

\begin{figure}
\begin{center}
\input{ABHfig5.tex}
\end{center}
\caption{For $g/m=4700$,
 plot of $\lg(\zeta/\zeta_{in})$ versus the number of e-folds of
 expansion in the reheat era.}   
\end{figure}

\begin{figure}
\begin{center}
\input{ABHfig6.tex}
\end{center}
\caption{For $g/m=4700$,
 logarithmic plot of energy densities of 
 $\varphi_1$ (A), $\varphi_0$ (B) and $\rho_0$ 
 versus the number of e-folds of expansion 
 in the reheat era.}   
\end{figure}

\end{document}